\documentclass[journal]{IEEEtran}

\usepackage{graphicx}
\usepackage{color}

\usepackage{subfigure}
\usepackage{multicol}
\usepackage{amssymb}
\usepackage{amsfonts}
\usepackage{amsmath}
\usepackage{psfrag}
\usepackage{url}
\usepackage{stfloats}
\usepackage{cite}
\usepackage{array}
\newtheorem{remark}{\bf Remark}
\newtheorem{theorem}{\bf Theorem}

\newtheorem{lemma}{\bf Lemma}

\begin{document}
\title{When reputation enforces evolutionary cooperation in unreliable MANETs}
\author{Changbing Tang, Ang Li, and Xiang Li, \IEEEmembership{Senior Member, IEEE}

\thanks{This work was partly supported by the National Key Basic Research and Development Program (Grant No. 2010CB731403),
the National Natural Science Foundation (No. 61273223),
the Research Fund for the Doctoral Program of Higher Education (No. 20120071110029)
and the Key Project of National Social Science Fund (No. 12\&ZD18) of China.}

\thanks{C. Tang is with the College of Mathematics, Physics and Information Engineering, Zhejiang Normal University, Jinhua 321004, China,
and also with the Adaptive Networks and Control Lab, Department of Electronic Engineering Fudan University, Shanghai 200433, China (e-mail: changbing1437@gmail.com).

A. Li and X. Li are with the Adaptive Networks and Control Lab, Department of Electronic Engineering Fudan University, Shanghai 200433, China (e-mail: 12210720031@fudan.edu.cn; lix@fudan.edu.cn). All correspondence should be addressed to X. Li.}
}

\maketitle

\begin{abstract}
In self-organized mobile ad hoc networks (MANETs), network functions rely on cooperation of self-interested nodes,
where a challenge is to enforce their mutual cooperation.
In this paper, we study cooperative packet forwarding in a one-hop unreliable channel which results from loss of packets and noisy observation of transmissions.
We propose an indirect reciprocity framework based on evolutionary game theory, and enforce cooperation of packet forwarding strategies in both structured and unstructured MANETs.
Furthermore, we analyze the evolutionary dynamics of cooperative strategies, and derive the threshold of benefit-to-cost ratio to guarantee the convergence of cooperation.
The numerical simulations verify that the proposed evolutionary game theoretic solution enforces cooperation when the benefit-to-cost ratio of the altruistic exceeds the critical condition.
In addition, the network throughput performance of our proposed strategy in structured MANETs is measured, which is in close agreement with that of the full cooperative strategy.
\end{abstract}


\section{Introduction}
\IEEEPARstart{I}{n} self-organized and distributed mobile ad hoc networks (MANETs) where each user belongs its own authority, the proper function of networks relies on cooperation of users (or nodes interchangeably).
Among all cooperative behaviors, packet forwarding enlarges the network coverage beyond one-hop transmission, which is particularly important in network formation.
However, fully cooperative behaviors cannot be taken for granted.
The nodes usually belong to different authorities, or work towards different goals, and pursue individual utilities \cite{Anderegg, Srinivasan1}.
A node incurs certain costs (power or money) in packet forwarding, which does not necessarily benefit itself.
As a result, the rational nodes maybe unwilling to participate in the forwarding, which damages the network performance \cite{Altman1, Yu}.
Hence, a challenging problem is to develop incentive mechanisms to encourage cooperative packet forwarding among selfish nodes to ensuring the proper functionalities of MANETs.

In recent years, two approaches have been proposed to steer users towards common network services.
One approach is to use the price mechanism to enforce cooperation \cite{Buttyan2, Crowcroft, Zhong, Janzadeh}.
A selfish node can gain payment from another node if the former helps forward the packets for the latter.
The payments can be money, or similar objects of value.
In \cite{Buttyan2}, Butty$\acute{a}$n and Hubaux introduced a concept of ``virtual currency" to implement the reward of users, participating in packet forwarding.
In \cite{Crowcroft}, Crowcroft et al. considered how to determine the price for the forwarding services to discourage the selfish behaviors in MANETs.
Authors in \cite{Zhong, Janzadeh} developed pricing schemes, cheat-proof schemes, and security of payment systems, which enforce collaborative users for forwarding packet in wireless networks.
The other approach is based on the reputation mechanism \cite{Michiardi1, Heq, Balakrishnan, Jaramillo, Refaei}, in which a node's forwarding decision depends on the history strategies of other nodes.
For instance, a reputation system, namely CORE, was proposed in \cite{Michiardi1} to enforce cooperation among selfish users by detecting misbehaving users.
He et al. presented a secure and objective reputation based incentive scheme \cite{Heq}.
Balakrishnan et al. let the source node choose the next hop node with sufficiently high reputation during the packet routing \cite{Balakrishnan}.
In addition, a distributed adaptive reputation mechanism and machine learning techniques were proposed in \cite{Jaramillo, Refaei} to provide the dynamical updating of reputation for cooperation.
Both approaches have pros and cons.
The pricing approach is simple in term of mechanism design, while it is difficult to implement in reality.
The reputation strategy, though involving more complicated reputation update, does not rely on a ``central bank'' to control the currency.

Meanwhile, a considerable amount of efforts have been devoted with game theory to analyzing how cooperation can be enforced \cite{Mejia1, Mejia2, Akkarajitsakul1, Duarte, Wangyf1, Wangyf2, Zeng}.
For example, in \cite{Srivastava2, Liz, Seredynski}, the authors applied game theory to analyze cooperation among selfish nodes,
and focused on the updating of individual' interaction strategies based on the behaviors of others in order to maximize their benefits.
In \cite{Wyu}, Yu and Liu proposed a game theoretic framework to analyze cooperation stimulation and security in autonomous mobile ad hoc networks.
In \cite{Aram}, the authors used game theory to optimize the allocation of resources in wireless networks and the packet routing in MANETs.
More close to the interest of this paper, F$\acute{e}$legyh$\acute{a}$zi et al. proposed a packet forwarding model in ad hoc networks based on game theory,
and derived the conditions under which cooperation yields the Nash Equilibrium \cite{Fele}.
Comprehensive review on this topic refer to Ref. \cite{Akkarajitsakul2, Khan1}.

One of the prominent properties of MANETs is unreliable channels between source and destination pairs.
A packet might be dropped due to link breakage or transmission errors even if other nodes are willing to forward \cite{Wyu, model3, model4}.
For example, in \cite{model3}, the authors proposed a feasible means based on the Bayesian formulation to achieve the equilibrium with the nodes' reputation in unreliable MANETs.
In their approach, a reputation model was employed, which requires the nodes to calculate their reputation about what actions the opponents have taken.
However, the computation complexity of updating reputation is high during the game evolution.
In fact, the model is co-evolutionary of reputation and strategy in nature, in which the performance of a node is the result of opponents reputation with other evolving entities in the system.
The higher reputation a node has, the more successfully the strategy of this node spreads in the system.
Besides, Ref. \cite{model4} addressed the enforcement cooperation of packet forwarding problem to discuss how cooperation can prevail over collusion using evolutionary game theory (EGT) in unreliable MANETs.
However, they did not consider the factor of network structure to the evolution of strategies.
Motivated by the aforemention challenge, an important question arises: how to develop a reputation strategy to enforce the cooperation in unreliable wireless networks, which is robust against packet loss and imperfect estimation of reputation?

In this work, we study cooperation enforcement of one-hop packet forwarding in unreliable MANETs.
Each node may act as a service provider, who has packets to transmit for certain destinations;
or act as a relay, who helps forward packets for the providers.
The provider will get certain benefits from the successful arrival of a packet, while the forwarding strategy of relay nodes will also incur certain costs.
Each node wants to maximize the chance of packet delivery with certain forwarding cost.
Hence, the packet forwarding can be described as a game, where players are selfish nodes, and the strategy of a player is to forward (F) or to discard (D) a packet.
During the process of forwarding packets, relay nodes may prefer not to participate in packet forwarding.
Our purpose is to develop a simple yet general reputation protocol to encourage selfish nodes to behave cooperatively.
More importantly, we explore how effective and robust a reputation strategy is.
An analytical framework based on evolutionary game theory is presented to study the dynamics and stability of player's strategies.
Our study differs from the state-of-art works in three aspects.
Firstly, we consider both cases of packet loss and error update of reputation caused by unreliable channels, while the existing works only consider the case of packet losses.
It is known that such two types of uncertainties have different impacts on cooperation enforcement.
Secondly, we propose a simple incentive protocol in the sense that the update of reputation does not require complicated iterations.
Finally, our work covers both unstructured and structured ad hoc systems.
In an unstructured system (USS), each node interacts with all other nodes.
In a structured system (SS), the interactions among nodes are characterized by a graph, where an edge denotes a pairwise interaction between two nodes.

Briefly, the main contribution of this paper can be summarized as follows:\\
$~~~ \bullet$ We model the packet forwarding process as a hidden action game with imperfect channels, and adopt EGT to capture the effectiveness and robustness of the proposed strategy in both USS and SS.\\
$~~~ \bullet$ We theoretically analyze the evolutionary dynamics of cooperative strategies, and derive the approximate threshold of benefit-to-cost ratio to guarantee the convergence of cooperation.\\
$~~~ \bullet$ We verify the cooperation enforcement with the indirect reciprocity mechanism in both USS and SS through extensive simulations.
Besides, the network throughput performance with channel loss probability and reputation updating error is provided.

The rest of this paper is organized as follows: Section II models the packet forwarding problem as a game, and introduces the advantage of evolutionary game theory in cooperation enforcement.
In Section III, we analyze the evolutionary dynamics of cooperative strategies, and obtain the thresholds of benefit-to-cost ratio in both unstructured and structured ad hoc systems.
Section IV gives numerical simulations to illustrate the proposed strategies.
Finally, section V concludes the whole paper.

\section{Modeling packet forwarding as games}
\subsection{Game model of packet forwarding}
We consider a self-organized MANET where pairwise nodes belong to different authorities.
Due to mobility and change of the environment, short-time interactions rather than long-time lasting associations between anonymous partners are dominant.
At each time slot, a fraction of players are chosen from the population to form pairs to forward packets \cite{Chaintreau2}.
Within each pair, players may either act as a service provider or a relay.
In addition, due to the constraint of communication range, the source of service provider cannot reach the destination directly.
We model such an one-hop packet forwarding as a peer-to-peer game between the pair of two nodes \cite{Seredynski, Fele, model3, model4}.
As shown in Fig. 1, there are two data sessions in the stage: (i) $S_{1}$ to $D_{1}$ through $S_{2}$, and (ii) $S_{2}$ to $D_{2}$ through $S_{1}$.
Each node $S_{1}$ (or $S_{2}$) chooses his strategy, $a_{i}$, from the strategy set $\mathbb{A}=\{F,D\}$, where $F$ and $D$ are packet forwarding and dropping, respectively.
At each time slot, node $S_{i}$ acts as a provider, and the other player acts as a relay; then the roles of two nodes switch.
That is to say, each node has $\frac{1}{2}$ chance to be a provider or a relay.

\begin{figure}[h]
\centering
\begin{tabular}{c}
\includegraphics[width=8.5cm]{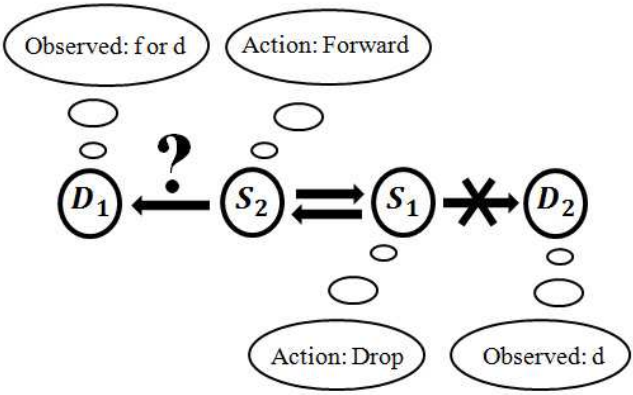}
\end{tabular}
\begin{center}
\caption{Two nodes' packet forwarding game model with unreliable channels.
At this stage, node $S_{2}$ forwards the packet for $S_{1}$, but the forwarding strategy might fail due to the channel noise,
thus the receiver $D_{1}$ of $S_{1}$ observes the signal of node $S_{2}$ is $f$ with probability $1-p_{e}$, or $d$ with probability $p_{e}$.
If $S_{1}$ drops the packet, the observed signal of node $S_{1}$ from $D_{2}$ is $d$.}
\end{center}
\end{figure}

If channels are reliable (loss free), the well-known Prisoner's Dilemma characterizes this scenario of packet forwarding \cite{model3, model4}.
For each node, when the packets are successfully delivered to the receiver (the destination of the packet forwarding), the provider will get a payoff, denoted as $b$.
Meanwhile, the forwarding effort of relay nodes will give rise to certain cost, denoted as $c$.
Thus, the payoff matrix between $F$ and $D$ is expressed as:
\begin{align}
\label{eq:0}
\bordermatrix{
  ~& \text{F} & \text{D} \cr
  \text{F} & b-c & -c \cr
 \text{D} & b & 0 \cr
}.
\end{align}

However, imperfect observation usually exists in such MANETs due to channel noise.
Although the nodes' strategies are hidden due to the channel, some traffic monitoring mechanisms are launched by each node to keep tracking of its neighbors' strategies \cite{Yu, Heq}.
Consider that the receiver of each node observes a private signal of the opponent's strategy from the set $\Theta=\{f,d\}$, where $f$ and $d$ are the observations of packet forwarding and dropping, respectively.
Since the node's observation is imperfect, the forwarding strategy $F$ of one node may be observed as $d$ by the other node due to link breakage or transmission errors.
Denote such channel loss probability as $p_{e}$.
For example, node $S_{2}$ forwards $S_{1}$'s packet to $D_{1}$, but the forwarding strategy might fail due to link breakage or transmission errors between $S_{2}$ and $D_{1}$ (see Fig. 1).

Besides, consider a reputation system in such MANETs, where each node is endowed with a binary reputation, denoted as good (G) or bad (B).
Based on the observation set $\Theta$, the reputation of a node evolves accordingly, which determines the strategy of other nodes adopted.
In some cases, the traffic monitoring mechanism of reputation collection can be unreliable, leading occasionally to false reports \cite{Seredynski}.
Thus, the reputation system must be fault tolerant.
In our model, this uncertainty is captured by parameter $\mu (0\leq\mu\leq 1)$, i.e., with probability $\mu$, an incorrect reputation is assigned; with probability $1-\mu$, a correct reputation is assigned.

\subsection{Why is evolutionary game theory?}
Consider the packet forwarding problem as a static game with noisy channels.
During the packet forwarding process, each node adapts its transmission probability, which depends on other nodes' strategies to maximize their own utility.
In the game, players are independent decision makers, and strive to maximize their own payoffs.
This similarity leads to a strong mapping between game theory components and elements of packet forwarding in MANETs \cite{Srivastava2}.

\textbf{Definition 1}:
A packet forwarding game $(\mathbb{G})$ with noisy channels is a quadruple $(I,\mathbb{A},\Theta,U)$, where\\
\textit{(I)} $I= 1, 2, \cdots, n$ denotes the set of players.\\
\textit{(II)} $\mathbb{A}=\times_{i\in I a_{i}}$ is a joint strategy set, $a_{i}\in\{F,D\}$ is the strategy of player $i$, $a_{-i}\in\{F,D\}$ is the strategy of the $i$th player's opponent.\\
\textit{(III)} $\Theta$ is the space of observed signals.
For every strategy $a_{i}$ that player $i$ takes, it observes a signal $\theta_{i}\in\Theta$.\\
\textit{(IV)} $U$ presents the realized payoffs.
For player $i$, its expected payoff is given by $u_{i}(a)=\sum_{\theta_{i}\in\Theta}\bar{u}_{i}(a_{i},a_{-i},\theta_{i})\cdot Prob(\theta_{i}\mid a_{-i})$, where $\bar{u}_{i}$ is the $i$th player's payoff.
\smallbreak

The outcome of a single static game can be characterized by the well-known Nash Equilibrium (NE), which is the strategy profile such that no player has a unilateral inventive to deviate and play another strategy.
However, the strategy profile which is beneficial for a given player might not always be beneficial for the whole system.
The question remains whether such strategies would eventually lead to a global cooperation or not.
The EGT is a nature-inspired approach to game theory, which reflects the dynamical evolution of strategies through pairwise interactions.
Therefore, the EGT is a suitable tool for analyzing the problem of cooperation in MANETs \cite{Seredynski}.

Correspondingly, the concept of Evolutionarily Stable Strategy (ESS) is central in evolutionary games.
Suppose that the whole population uses strategy $q$, and a small fraction $\epsilon$ (called ``mutations") adopts strategy $p$ $(p\neq q)$.

\textbf{Definition 2} \cite{Hofbauer}:
Evolutionary forces are expected to select $q$ against $p$ if:
\begin{equation}
\label{eq:1}
\begin{array}{c}
U(q,\epsilon p+(1-\epsilon)q)>U(p,\epsilon p+(1-\epsilon)q).
\end{array}
\end{equation}
Strategy $q$ is said to be the ESS if for every $p\neq q$, there exists some $\hat{\epsilon}_{y}> 0$ such that Eq. $\eqref{eq:1}$ holds for all $\epsilon\in(0,\hat{\epsilon}_{y})$.

The definition of ESS is stronger than that of NE, as the former is robust against a deviation of the whole population, while the latter only concerns the deviations of a single player.
Although ESS has been originally defined in biological systems, it is highly relevant to engineering as well \cite{Vincent1, Yang, Altman3, Khan2}.
In addition, there are two advantages within the framework of evolutionary games \cite{Tembine}: 1) It allows us to identify robustness against deviations of more than one player,
and 2) it allows us to apply the generic convergence theory of evolutionary game dynamics and
stability to capture the effectiveness and robustness of the proposed strategy.

To investigate the cooperative character of proposed strategies, we adopt another evolutionarily stable definition in this paper.

\textbf{Definition 3} \cite{Ohtsuki1}:
A strategy is a Cooperative Evolutionarily Stable Strategy (CESS) if and only if it satisfies the following two criteria: \\
\textit{(I)} Cooperativity (CO): more than the half of all game interactions are cooperative. \\
\textit{(II)} Evolutionary stability (ES): strategy $q$ is evolutionarily stable against any other strategy $p$ $(p\neq q)$.

\section{Cooperation enforcement in unreliable MANETs with indirect reciprocity}

\begin{figure}[h]
\centering
\begin{tabular}{c}
\includegraphics[width=8cm]{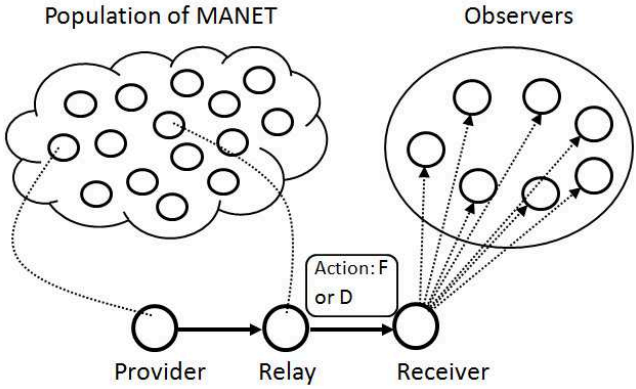}
\end{tabular}
\begin{center}
\caption{The illustration of the indirect reciprocity mechanism.}
\end{center}
\end{figure}

Indirect reciprocity is a powerful mechanism for the evolution of cooperation, and has recently drawn a lot of attentions \cite{Ohtsuki1, Nowak2, Cheny}.
The essential concept of indirect reciprocity is ``I help you not because you have helped me but because you have helped others"\cite{Nowak2}.
Therefore, a key problem in the indirect reciprocity mechanism is the establishment of reputation, which is the evaluation of the history of players' actions.
As shown in Fig. 2, within every interaction, a pair of provider and relay are randomly sampled from the MANET.
Then, the relay will forward or drop the packets of the provider to the receiver according to the provider's reputation.
After the transmission, the relay's reputation will be updated based on the observed signal of receiver.
Finally, the relay's reputation is propagated to the whole population from the receiver and the observers through a noisy gossip channel.
Since we consider the scene of two nodes' packet forwarding game as shown in Fig. 1,
the roles of provider and relay are switched after the aforementioned process of packet forwarding is finished.

Generally, helping someone establishes a good reputation, and will be rewarded by others.
In this paper, we adopt the reputation updating rule of indirect reciprocity in \cite{Ohtsuki1}, i.e., the reputation of relay is updated according to the following rule:
\begin{equation}
\label{eq:3}
\begin{array}{ccc}
  ~& \text{G} & \text{B} \\
  \text{F} & G & G \\
  \text{D} & B & G
\end{array}
\end{equation}
where a relay who takes the choice $X(X\in\{F,D\})$ towards a provider with reputation $R$ $(R\in\{G,B\})$ will be assigned a new reputation $R^{'}(R;X)$ $(R^{'}\in\{G,B\})$.
Here, we adopt the reputation updating such that cooperation leads to a good reputation, whereas defection leads to a bad reputation unless the opponent is a bad player.

Based on the above reputation system, each player will select a new strategy, $\tilde{a}_{i}$, which depends on the provider's reputation.
Specifically, a player with $\tilde{a}_{i}$ takes strategy $s(G)$ $(s(G)\in\{F,D\})$ for a good provider, and strategy $s(B)$ $(s(B)\in\{F,D\})$ for a bad one.
Thus, the new strategy set of player, $\tilde{\mathbb{A}}$, has $2^{2}=4$ possible elements: $\tilde{\mathbb{A}}=\{s(G)s(B)|FF, FD, DF, DD\}$.
For example, $FD$ means that taking strategy $F$ towards a good provider and strategy $D$ towards a bad one.
In this paper, we only consider three of these strategies, i.e., $FF$, $FD$ and $DD$, since strategy $DF$ is illogical in practice.
Denote $x_{1}$, $x_{2}$, $x_{3}$ as the frequencies of strategy $FF$, $FD$, $DD$,
and $r_{1}$, $r_{2}$, $r_{3}$ as the frequencies of players with good reputation among $FF$, $FD$, $DD$ players, respectively.
Therefore, the summed frequency of players with good reputation in the entire system is $r = x_{1}r_{1} + x_{2}r_{2} + x_{3}r_{3}$.

Consider the fast reputation updating that all players update their strategies only after they actually conceive the payoff differences from different strategies.
Thus, during the evolution of strategy and reputation, the reputation quickly converges to a stable state.
The following lemma gives the stable reputation distribution of the system.
\smallbreak

\begin{lemma}
Given the fast reputation updating, the players' reputation distribution converges to a stable state, i.e., $\exists t_{N}$,
when $t\rightarrow t_{N}$, $lim_{t\rightarrow t_{N}} r_{m}=r_{m}^{*}$ ($m=1,2,3$ denotes the strategy $FF$, $FD$, $DD$, respectively), where
\begin{equation}
\label{eq:50}
\left\{%
\begin{array}{lll}
r^{*}_{1}=1-\mu\\
r^{*}_{2}=1-\mu\\
r^{*}_{3}=(1-\mu)[1-\frac{1-2\mu}{1+(1-2\mu)x_{3}}].
\end{array}%
\right.
\end{equation}
\end{lemma}
\begin{proof}
See Appendix $B$.
\end{proof}

\subsection{Cooperation enforcement in an USS}
Consider an unstructured ad hoc system with $N$ players.
At each time slot, a fraction of players are chosen from the population to form pairs, and in each pair the players need to help each other to forward packets to the destination.
Firstly, one player acts as a relay, and the other player acts as a service provider; then the roles of two nodes switch.
In the game level, the relay has two choices: Forward (F) or Drop (D), and the service provider does nothing.
Calculating $u(a)$ for different strategy pairs, we have the strategic form of the packet forwarding game, where players' payoff matrix $M_{1}$ is given by:
\begin{align}
\label{eq:2}
\bordermatrix{
  ~& \text{Forward} & \text{Drop} \cr
  \text{Forward} & b(1-p_{e})-c& -c \cr
  \text{Drop} & b(1-p_{e})& 0 \cr
}.
\end{align}
Note that the players' payoff matrix $M_{1}$ under indirect reciprocity can be obtained from $\eqref{eq:0}$ with the possibility of channel loss.
Specifically, the gain of a provider is $b(1-p_{e})$ when the packets are successfully delivered to the destination, and the cost of a relay with forwarding strategy is $c$.

Given a stable reputation distribution, the expected payoff of a strategy can be calculated.
For a $FF$ player, he has $\frac{1}{2}$ chance to be a relay and cooperate with cost $c$.
With $\frac{1}{2}$ chance being a provider, he meets a $FF$, $FD$ and $DD$ player with probability $x_{1}$, $x_{2}$ and $x_{3}$, and is expected to get the gain of $b(1-p_{e})$, $b(1-p_{e})(1-\mu)$ and $0$, respectively.
Therefore, the expected payoff of all three strategies are
\begin{equation}
\label{eq:4}
\left\{%
\begin{array}{lll}
P_{1}=\frac{1}{2}(-c)+\frac{1}{2}[b(1-p_{e})x_{1}+b(1-p_{e})r_{1}x_{2}]\\
P_{2}=\frac{1}{2}r(-c)+\frac{1}{2}[b(1-p_{e})x_{1}+b(1-p_{e})r_{2}x_{2}]\\
P_{3}=\frac{1}{2}(0)+\frac{1}{2}[b(1-p_{e})x_{1}+b(1-p_{e})r_{3}x_{2}],
\end{array}%
\right.
\end{equation}
where $P_{1}$, $P_{2}$, and $P_{3}$ are the expected payoffs of strategy $FF$, $FD$, and $DD$, respectively.

\begin{figure*}
\begin{center}
\includegraphics[width=11cm]{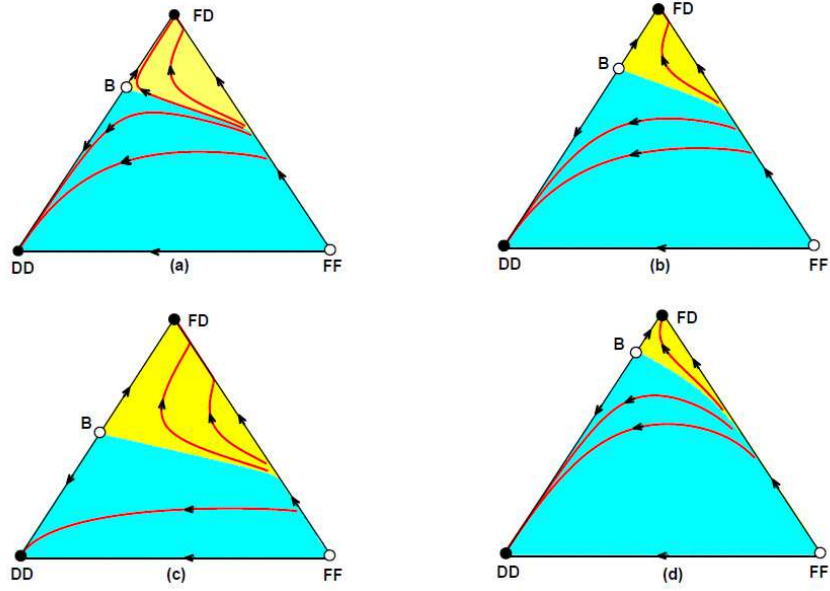}
\caption{Phase portrait of Eq. $\eqref{eq:5}$.
Each vertex represents a state with players taking the same strategy, such that point $FD$ represent the state $(0, 1, 0)$ in which all players taking $FD$ strategy. The upper yellow part is the attraction basin of $FD$-type CESS, and the lower light blue part is the attraction basin of $DD$-type CESS.
The separatrix line is the stable manifold of saddle point $B$. We set the system parameters as $\beta=10$, and (a) $b=3$, $c=2$, $\mu=0.01$, $p_{e}=0.01$; (b) $b=3$, $c=2$, $\mu=0.01$, $p_{e}=0.08$; (c) $b=4$, $c=2$, $\mu=0.01$, $p_{e}=0.01$; (d) $b=3$, $c=2$, $\mu=0.1$, $p_{e}=0.01$.}
\end{center}
\end{figure*}

Adopt the Fermi update dynamics \cite{Blume} to describe the strategy evolution, where players update their strategies to get a higher payoff by imitating the better strategy.
At each time step, two players ($i$ and $j$) are randomly chosen.
Then, the strategy of player $i$ will replace that of player $j$ with probability
\begin{equation}
\label{eq:40}
\begin{array}{c}
p=[1+e^{-\beta(P_{i}-P_{j})}]^{-1},
\end{array}
\end{equation}
where $P_{i}$ is the payoff of player $i$, and $\beta\geq0$ is the intensity of imitation.
From Eq. $\eqref{eq:40}$, we know that the higher payoff of a player's strategy has, the more possible the strategy is to imitate by other players.
So the payoff of a strategy can be interpreted as its fitness, and strategies with higher fitness have more chance to reproduce.

For the pairwise comparison process, the evolutionary dynamics can be approximated by \cite{Traulsen2}: $\dot{x}_{m}=\frac{\beta}{2}x_{m}(P_{m}-\bar{P})$ $(m=1,2,3)$,
where $\bar{P}=\Sigma_{m=1}^{3}x_{m}P_{m}$ is the average payoff of three strategies.
Define $\hat{P}_{m}=P_{m}-P_{3}$, and $\tilde{P}=\Sigma_{m=1}^{3}x_{m}\hat{P}_{m}$.
We get the transformed dynamics of strategies frequency as
\begin{equation}
\label{eq:5}
\left\{%
\begin{array}{llll}
\dot{x}_{1}=\frac{\beta}{2}x_{1}(\hat{P}_{1}-\tilde{P})=-cx_{1}+cx_{1}^{2}\\
~~~~~+\frac{[(1-2\mu)b(1-p_{e})+c]x_{1}x_{2}-(1-2\mu)b(1-p_{e})x_{1}x_{2}(x_{1}+x_{2})}{2-\frac{1-2\mu}{1-\mu}(x_{1}+x_{2})}\\
\dot{x}_{2}=\frac{\beta}{2}x_{2}(\hat{P}_{2}-\tilde{P})=cx_{1}x_{2}\\
~~~~~+\frac{-cx_{2}+[(1-2\mu)b(1-p_{e})+c]x_{2}^{2}-(1-2\mu)b(1-p_{e})x_{2}^{2}(x_{1}+x_{2})}{2-\frac{1-2\mu}{1-\mu}(x_{1}+x_{2})}
\end{array}%
\right.
\end{equation}

Note that Eq. $\eqref{eq:5}$ is defined on simplex $S_{3}=\{(x_{1},x_{2},x_{3})|x_{1}+x_{2}+x_{3}=1,x_{m}\geq 0\}$,
each corner of the simplex is an equilibrium of the dynamics corresponding to a monomorphic state.

\begin{theorem}
Given a stable reputation distribution satisfying Eq. $\eqref{eq:50}$, if the benefit-to-cost ratio of the altruistic $\frac{b}{c}>\frac{1}{(1-2\mu)(1-p_{e})}$, strategy $FD$ is a CESS.
\end{theorem}
\smallbreak

\begin{proof}
Note that state $(x_{1}, x_{2}, x_{3})=(0, 1, 0)$ is the corner of $S_{3}$, it is an equilibrium of Eq. $\eqref{eq:5}$.
For this equilibrium, the Jacobian matrix $J$ of Eq. \eqref{eq:5} has the form
\begin{equation*}
\centering
\mathbf{J}|_{x_{1}=0, x_{2}=1}=\left(%
\begin{array}{ccc}
  J_{11} & J_{12} \\
 J_{21} & J_{22} \\
\end{array}%
\right),
\end{equation*}
where $J_{11}=-\mu c$, $J_{12}=0$,
$J_{21}=-c-(1-\mu)(1-2\mu)b(1-p_{e})$, $J_{22}=(1-\mu)[c-(1-2\mu)b(1-p_{e})]$.
The corresponding eigenvalues of the Jacobian matrix $J$ at $x_{1}=0$, $x_{2}=1$ are $\lambda_{1}=-\mu c$, $\lambda_{2}=(1-\mu)[c-(1-2\mu)b(1-p_{e})]$.
Since $\mu>0$, $c>0$, so $\lambda_{1}<0$.
When $\frac{b}{c}>\frac{1}{(1-2\mu)(1-p_{e})}$, i.e. $c-(1-2\mu)b(1-p_{e})<0$, $\lambda_{2}<0$, the state $(0, 1, 0)$ is stable, which arrives that strategy $FD$ is an ES.

In order to get a good reputation following Eq. $\eqref{eq:3}$, one has to cooperate with good providers, and may take an arbitrary strategy ($F$ or $D$) against bad providers.
However, the best choice with bad providers is to drop since adopting strategy $D$ has no cost.
Thus, the best choice of a player is $s(G)s(B)=FD$.
In such an equilibrium, more than the half of interactions are cooperation.
With Definition 3, we know that strategy $FD$ is a cooperative evolutionarily stable strategy (CESS).
\end{proof}
\smallbreak

\begin{remark}
\textit{(I)} Note that $1-2\mu$ represents the system discrimination of the player reputation updating ($0 <\mu <1/2$). Denote $q=1-2\mu$.
When $\mu=0$, $q=1$, the updating of reputation is in error-free; when $\mu=1$, $q=1/2$, system \eqref{eq:5} can not distinguish players with good reputation and bad reputation.
Thus, $q$ reflects the discrimination ability of system \eqref{eq:5}.

\textit{(II)} Transforming the condition in Theorem 1, we get $\frac{c}{b}<q(1-p_{e})$.
This implies that if the multiplier of $q$ and $1-p_{e}$ exceeds the cost-to-benefit ratio $\frac{c}{b}$, $FD$ becomes a CESS.
Therefore, the indirect reciprocity mechanism enforces cooperation under appropriate parameters in this unstructured ad hoc system.
\end{remark}

In Fig. 3, we illustrate the phase portrait of Eq. $\eqref{eq:5}$ with different parameters, where strategies $FD$ and $DD$ are evolutionary stable,
while strategy $FF$ is unstable.
Decreasing the probability of transmission error $p_{e}$ and reputation updating error $\mu$ (or increasing benefit $b$) will enlarge the attraction basin of $FD$-type CESS,
i.e., it is easier for cooperation thrives when $p_{e}$, $\mu$ are small and $b$ is large.
Here, the attraction basin of a strategy are the sets of all initial strategy distributions in a feasible domain that converge to the CESS.
Therefore, given appropriate system parameters to satisfy the conditions of Theorem $1$ and the strategy distributions in the attraction basin of $FD$-type CESS, cooperation of packet forwarding in MANETs can be enforced with the indirect reciprocity mechanism.

\subsection{Cooperation enforcement in a SS}
So far, we have provided a method to enforce cooperation in an unstructured ad hoc system.
However, the interactions of real MANETs are often restricted to a small group, and every player has its own forwarding domain.
In this subsection, we focus on how cooperation can be enforced among the players in a SS, where the players' interactions reflect the structure of the MANET.

Consider that there are $N$ players who are located at nodes in a MANET, and the average degree is $L$.
Each player adopts one strategy of three types, $FF$, $FD$ or $DD$.
Consider the packet forwarding game among $FF$, $FD$ and $DD$.
Then, the payoff matrix $M_{2}$ between two players is:
\begin{align}
\label{eq:6}
\bordermatrix{
  ~& FF & FD & DD \cr
  FF & b(1-p_{e})-c& -c+r_{1}(1-p_{e})b & -c \cr
  FD & r(-c)+b(1-p_{e})& r(-c)+r_{2}b(1-p_{e})& r(-c)\cr
  DD & b(1-p_{e}) & r_{3}b(1-p_{e}) & 0\cr
}.
\end{align}

\begin{figure}[ht]
\centering
\begin{tabular}{c}
\includegraphics[width=7.2cm]{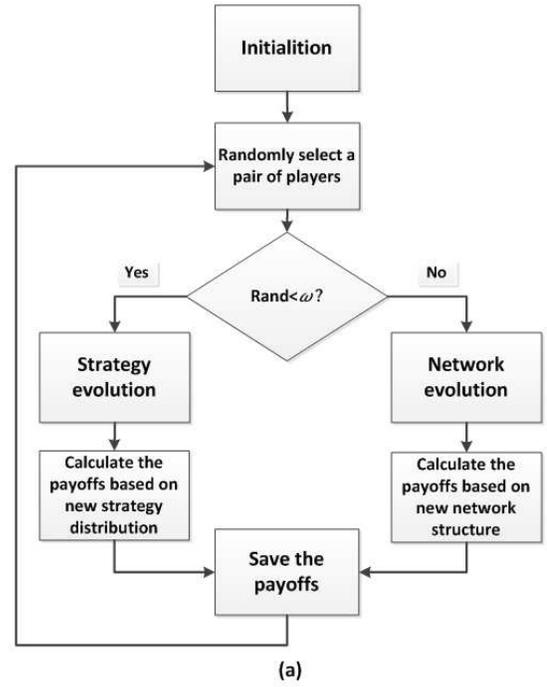}\vspace{0.5cm} \\
\includegraphics[width=7.2cm]{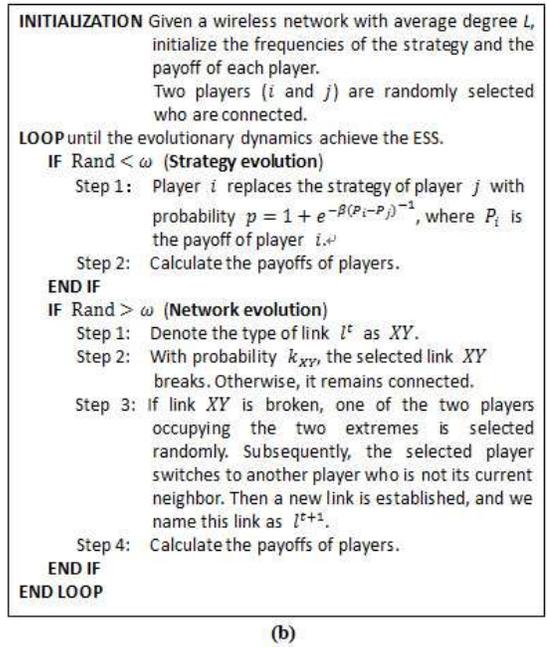}
\end{tabular}
\begin{center}
\caption{(a) Flowchart of the co-evolution of strategy and network structure.
(b) Schematic operation of the co-evolution of strategy and network structure.}
\end{center}
\end{figure}

In fact, one feature of MANETs lies that the topology is dynamical, which is expressed in terms of mobility and connectivity of the players \cite{Fischer}.
Here, we introduce co-evolutionary rules of strategy and network topology.
At each time step, the strategic evolution happens with probability $\omega$;
otherwise, the topological evolution of the network occurs with probability $1-\omega$.
Note that probability $\omega$ denotes the dynamical time-scale of these two updating rules.
To characterize the dynamics of network structure where players leave or break interactions when they dissatisfy with the current situation,
we denote $k_{XY}$ ($X,Y\in\{FF,FD, DD\}$) as the probability with which an $XY$-type link breaks.

We summarize the co-evolution of strategy-updating and structure-switching as follows:

\emph{Strategy updating}: We adopt the same pairwise comparison rule on networks as that in the previous subsection,
i.e., the strategy of player $i$ replaces that of player $j$ with probability $p=[1+e^{-\beta(P_{i}-P_{j})}]^{-1}$.

\emph{Network switching}: Each link is assigned a number $l\in\{1,2,\cdots, H\}$ as its label, where $H=\frac{LN}{2}$ is the total number of links.
Denote $E_{3}=\{XY\mid X,Y = FF,FD, or~DD \}$ as the collection of all possible links.
Briefly, the network evolves as:\\
\textit{Step~1}: At time step $t$, a link $l^{t}$ of type $XY\in E_{3}$ is selected from the network at random.\\
\textit{Step~2}: With probability $1-k_{XY}$, the selected link $l^{t}$ remains unchanged, denoted as $l^{t+1}=l^{t}$.
With probability $k_{XY}$, the selected link breaks.\\
\textit{Step~3}: If the link is indeed broken, one of the two players occupying the end points of the link is randomly selected.
The selected player switches to another player who is not its current neighbor.
Then a new link is established, and we name this link as $l^{t+1}$.

\begin{figure*}
\begin{center}
\includegraphics[width=10.5cm]{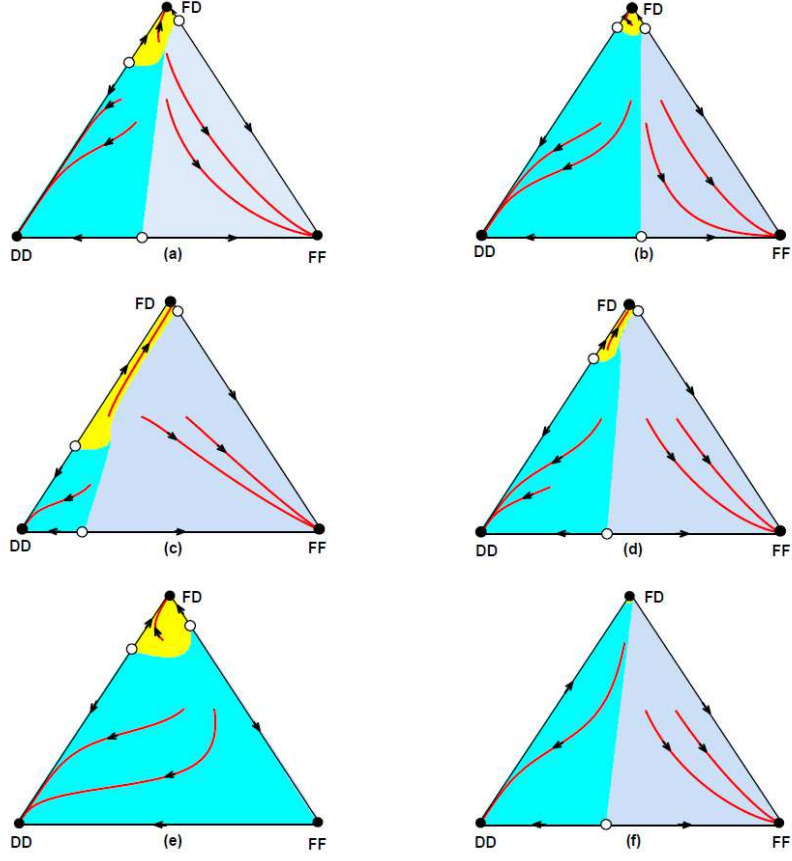}
\caption{Phase portrait of Eq. $\eqref{eq:10}$.
Each vertex represents a state with players taking the same strategy.
The left upper yellow part is the attraction basin of $FD$-type CESS, the left lower light blue part is the attraction basin of $DD$-type CESS,
and the right lower gray part is the attraction basin of $FF$-type CESS.
The separatrix line is the stable manifold of saddle point.
We set the system parameters as $\beta=10$, $L=4$, $k_{12}=0.25$, $k_{13}=0.3$, $k_{23}=0.95$, and (a) $p_{e}=0.01$, $b=3$, $c=2$, $\mu=0.1$, $k_{11}=0.05$, $k_{22}=0.25$; (b) $p_{e}=0.08$, $b=3$, $c=2$, $\mu=0.1$, $k_{11}=0.05$, $k_{22}=0.25$; (c) $b=4$, $c=2$, $\mu=0.1$, $p_{e}=0.01$, $k_{11}=0.05$, $k_{22}=0.25$; (d) $\mu=0.01$, $b=3$, $c=2$, $p_{e}=0.01$, $k_{11}=0.05$, $k_{22}=0.25$; (e) $k_{11}=0.1$, $b=3$, $c=2$, $\mu=0.1$, $p_{e}=0.01$, $k_{22}=0.25$; (f) $k_{22}=0.35$, $b=3$, $c=2$, $\mu=0.1$, $p_{e}=0.01$, $k_{11}=0.05$.}
\end{center}
\end{figure*}

Note that $k_{XY}$ is time-invariant, and it reflects the network effect of linking dynamics.
Furthermore, the inverse of $k_{XY}$ can be regarded as the average duration time between strategies $X$ and $Y$.
In other words, $k_{XY}$ is a measurement of the duration of $XY$ link.
The implementation of the co-evolution of strategy and structure scheme is depicted as shown in Fig. 4.

Denote the type of a selected link $l^{t}$ as $T(l^{t})$, where $T(l^{t})\in$ $\{FF-FF$, $FF-FD$, $FF-DD$, $FD-FD$, $FD-DD$ and $DD-DD\}$.
The dynamics of $T(l^{t})$ can be described as a Markov chain with transition matrix $Q=[Q_{(XY)(ZW)}]_{3\times 3}$,
and each entry $Q_{(XY)(ZW)}$ indicates the transition probability that a $XY$-type link transforms to a $ZW$-type link at each time step.
Note that such a Markov chain is irreducible and aperiodic, and it leads to the stationary distribution \cite{Wu2}
\begin{equation}
\label{eq:7}
\begin{array}{l}
\pi_{XY}=a(x)(2-\delta_{XY})x_{X}x_{Y}/k_{XY}, X, Y\in E_{3},
\end{array}
\end{equation}
where $\delta_{XY}$ indicates the Kronecker delta, $x=(x_{1}, x_{2}, x_{3})^{T}$ is the vector frequency of strategy $FF$, $FD$ and $DD$, and $a(x)=[(x_{1}^{2}/k_{11})+(x_{2}^{2}/k_{22})+(x_{3}^{2}/k_{33})+(2x_{1}x_{2}/k_{12})+(2x_{1}x_{3}/k_{13})+(2x_{2}x_{3}/k_{23})]^{-1}$ is the normalization factor.
We then have the average number of $XY$-type links $N_{XY}$, i.e., $E(N_{XY})=H\pi_{XY}$.

When $\omega\ll 1$, the players in a MANET are much more reluctant to change strategies than to adjust their partnerships \cite{Tang1, Wu2}.
In this case, the structure evolution of the network obeys a stationary distribution which is described by Eq. $\eqref{eq:7}$ when the strategy evolution occurs.
Moreover, the average fitness of a player is determined by the stationary distribution of the linking dynamics.
Therefore, the average fitness functions of strategy $m$ ($m=1,2,3$) is
\begin{equation}
\label{eq:8}
\begin{array}{ll}
f_{m}&=\sum_{m^{'}=1}^{3}E(N_{mm^{'}})M_{2}(1+\delta_{mm^{'}})/(Nx_{m})\\
&=La(x)(M_{2}^{'}x)_{m},
\end{array}
\end{equation}
where $M_{2}^{'}$ is given by
\begin{align}
\label{eq:9}
\bordermatrix{
  ~& FF & FD & DD \cr
  FF & \frac{b(1-p_{e})-c}{k_{11}}& \frac{-c+r_{1}(1-p_{e})b}{k_{12}} & \frac{-c}{k_{13}} \cr
  FD & \frac{r(-c)+b(1-p_{e})}{k_{12}}& \frac{r(-c)+r_{2}b(1-p_{e})}{k_{22}}& \frac{r(-c)}{k_{23}}\cr
  DD & \frac{b(1-p_{e})}{k_{13}} & \frac{r_{3}b(1-p_{e})}{k_{23}} & 0\cr
}.
\end{align}
Eq. \eqref{eq:8} suggests that the co-evolution of strategy and network structure introduces a simple transformation of payoff matrix $M_{2}$.
\smallbreak

\begin{theorem}
Given a stable reputation distribution satisfying Eq. $\eqref{eq:50}$, the co-evolution of strategy and network structure with indirect reciprocity leads that:\\
\textit{(I)} Strategy $FF$ is a CESS, if the benefit-to-cost ratio of the altruistic $\frac{b}{c}>max\{\frac{k_{12}-k_{11}(1-\mu)}{(k_{12}-k_{11})(1-p_{e})}, \frac{k_{13}}{(k_{12}-k_{11})(1-p_{e})}\}$;\\
\textit{(II)} Strategy $FD$ is a CESS, if the benefit-to-cost ratio of the altruistic $\frac{b}{c}$ satisfies:  $\frac{k_{23}}{(k_{23}-k_{22})(1-p_{e})}<\frac{b}{c}<\frac{k_{22}-k_{12}(1-\mu)}{(k_{22}-k_{12})(1-\mu)(1-p_{e})}$.
\end{theorem}
\smallbreak

\begin{proof}
\textit{(I)} From the first column of $M_{2}^{'}$, when $\frac{b(1-p_{e})-c}{k_{11}}>\frac{r(-c)+b(1-p_{e})}{k_{12}}$, and $\frac{b(1-p_{e})-c}{k_{11}}>\frac{b(1-p_{e})}{k_{13}}$, i.e., $\frac{b}{c}>max\{\frac{k_{12}-k_{11}(1-\mu)}{(k_{12}-k_{11})(1-p_{e})}, \frac{k_{13}}{(k_{12}-k_{11})(1-p_{e})}\}$, $FF$ becomes a CESS, which means that if all players take $FF$ strategy, the best choice for one player is to take strategy $FF$.

\textit{(II)} Similar to \textit{(I)}, when $\frac{r(-c)+r_{2}b(1-p_{e})}{k_{22}}>\frac{-c+r_{1}(1-p_{e})b}{k_{12}}$, and $\frac{r(-c)+r_{2}b(1-p_{e})}{k_{22}}>\frac{r_{3}b(1-p_{e})}{k_{23}}$, i.e., $\frac{k_{23}}{(k_{23}-k_{22})(1-p_{e})}<\frac{b}{c}<\frac{k_{22}-k_{12}(1-\mu)}{(k_{22}-k_{12})(1-\mu)(1-p_{e})}$, $FD$ becomes a CESS.
In this case, when all players take $FD$ strategy, the best choice for one player is to take strategy $FD$.
\end{proof}
\smallbreak

\begin{remark}
\textit{(I)} Comparing to the case of USS, the critical value of the benefit-to-cost ratio of the altruistic in a structured system not only includes the system resolution $p_{e}$ and incorrect reputation assignment $\mu$, but also embeds the network effect of linking dynamics $k_{ij}$.

\textit{(II)} Obviously, $\frac{k_{23}}{(k_{23}-k_{22})(1-p_{e})}<\frac{2}{(1-2\mu)(1-p_{e})}$.
Therefore, the condition for strategy $FD$ becoming an ESS is easier in the structured system than that in the unstructured system.
Moreover, when the benefit-to-cost ratio of the altruistic exceeds the critical value, $\frac{b}{c}>max\{\frac{k_{12}-k_{11}(1-\mu)}{(k_{12}-k_{11})(1-p_{e})}, \frac{k_{13}}{(k_{12}-k_{11})(1-p_{e})}\}$, strategy $FF$ becomes an ESS.
This situation never happens in an unstructured system.
Thus, the effects of the co-evolutionary of strategy and structure scheme can enforce cooperation in MANETs with the indirect reciprocity mechanism.
\end{remark}

For a general imitation process, the strategy evolution can be approximated by high-dimensional stochastic differential equations \cite{Traulsen2}.
When $N\gg L$, each player has a very limited number of neighbors compared to the population size of a network,
and the stochastic term vanishes, which yields:
\begin{equation}
\label{eq:10}
\begin{array}{ll}
\dot{x}_{m}=\frac{\beta}{2}La(x)x_{m}((M_{2}^{'}x)_{m}-x^{T}M_{2}^{'}x).
\end{array}
\end{equation}

We illustrate the CESS of Eq. $\eqref{eq:10}$ by the phase portrait as shown in Fig. 5.
Compared with the unstructured case of Fig. 3, Fig. 5 tells that strategies $FF$, $FD$ and $DD$ all are evolutionary stable.
That is, given appropriate system parameters satisfying condition \textit{(I)} in Theorem $2$, and the strategy distributions in the attraction basin of $FF$-type CESS, the fully cooperation state can be obtained.
Note that fully cooperation strategy outperforms all other non-cooperative strategies, which leads to high performance of packet forwarding in MANETs.
Thus, the co-evolution of strategy and structure scheme with indirect reciprocity significantly promotes cooperation.

Similarly, Fig. 5 shows that decreasing the probability of transmission error $p_{e}$ and reputation updating error $\mu$ (or increasing the benefit $b$) enlarge the attraction basin of $FD$-type and $FF$-type CESS, i.e., cooperation is easier to thrive when $p_{e}$, $\mu$ are small and $b$ is large.
Besides, consider the effect of network structure, decreasing the probability of rewiring $k_{11}$ (decreasing the probability of rewiring $k_{22}$) can also enlarges the attraction basin of $FF$-type CESS (enlarges the attraction basin of $FD$-type CESS).

\section{Numerical Results}
\subsection{Simulation setup}
Our simulations aim at one-hop packet forwarding scenario (see Fig. 1) in MANETs, where the two-player packet forwarding game based on indirect reciprocity can be directly applied to.
In the USS, at each time slot, two nodes are randomly picked to form a pair, and forward packets for each other.
In the SS, we consider a dynamical MANET, and players are able to forward packets for each other only when they interact mutually.
Initially, players are situated on the nodes of a regular graph with degree $L=4$.
Subsequently, the network structure and strategy of players evolve with the co-evolution scheme as shown in Fig. 4.

In order to study the performance of our proposed EGT-based approach in USS and SS,
we study the cooperation level when all strategies converge to the $FD$-type CESS in the USS and the $FF$-type CESS in the SS, respectively.
We fix the payoff matrix as $b=4$, $c=2$, and $\omega=0.02$, $\beta=10$ in our simulations.
Besides, set network parameters $k_{11}=0.05$, $k_{12}=0.25$, $k_{13}=0.3$, $k_{22}=0.25$, and $k_{23}=0.95$,
which satisfies the conditions of Theorem 2 in the SS.

\subsection{Performance evaluation}
Contrastively, we compare the performance of our proposed EGT-based strategy in USS and SS with ``Full cooperation".
Full cooperation implies that a node will always forward other's packets unconditionally.
Such a full cooperation strategy is not implementable in autonomous MANETs,
but it can serve as a performance upper bound of the proposed strategy to measure the performance loss due to noisy and imperfect observation \cite{model3}.

Fig. 6 shows the average payoff for each of the nodes in USS and SS based on forwarding strategy, respectively.
In this setting, $p_{e}=0.01$ and $\mu=0.01$.
It presents that our proposed EGT-based indirect reciprocity approach in small noisy channels can enforce cooperation in not only the USS but also the SS.
The average payoffs of both cases are much closer to the upper bound than the lower bound when the loss and channel noise are small,
and mutual cooperation is enforced.
Furthermore, the average payoffs in the SS are higher than those of the USS, which implies that the co-evolution of strategy and structure with indirect reciprocity promote cooperation.

\begin{figure}[h!]
\centering
\begin{tabular}{c}
\includegraphics[width=8.2cm]{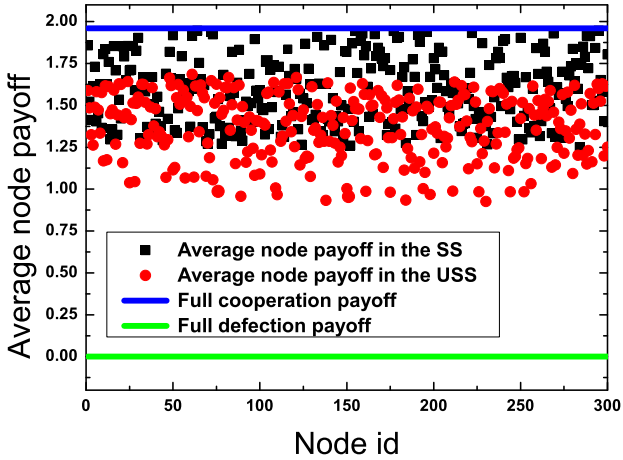}
\end{tabular}
\begin{center}
\caption{Average node payoff for the USS and the SS based on the EGT of indirect reciprocity approach at $p_{e}=0.01$, $\mu=0.01$.}
\end{center}
\end{figure}

\begin{figure}[h!]
\centering
\begin{tabular}{c}
\includegraphics[width=7.6cm]{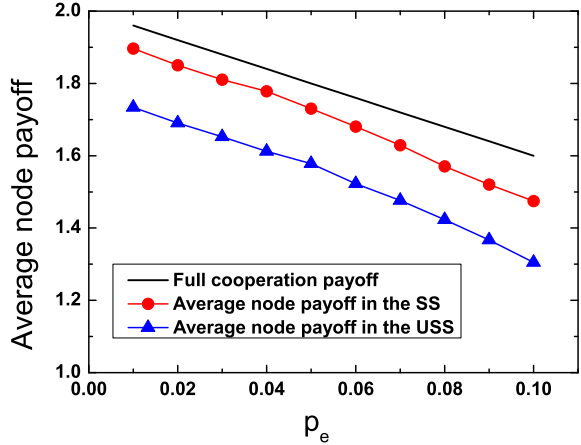}
\end{tabular}
\begin{center}
\caption{Average node payoff for the USS and the SS based on the EGT of indirect reciprocity approach with different $p_{e}$ at $\mu=0.01$.}
\end{center}
\end{figure}

Fig. 7 and Fig. 8 illustrate the average node payoffs in both USS and SS based on the EGT of indirect reciprocity approach with different channel loss probability $p_{e}$ and reputation updating error $\mu$.
In Fig. 7, reputation updating error is set as $\mu=0.01$, and in Fig. 8, channel loss probability is set as $p_{e}=0.01$.
The plots show that the average payoff drops when the channel becomes more unreliable (large $p_{e}$) and the reputation becomes more undistinguishable (large $\mu$).
Especially, with increasing the value of $\mu$, the difference of average node payoffs between ``Full cooperation" and the proposed strategy in the USS becomes large.
These observations also suggest that the performance in the SS is better than that in the USS.

\begin{figure}[h!]
\centering
\begin{tabular}{c}
\includegraphics[width=7.7cm]{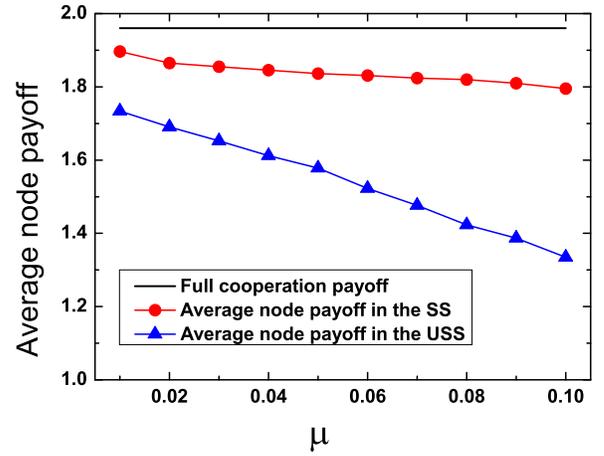}
\end{tabular}
\begin{center}
\caption{Average node payoff for the USS and the SS based on the EGT of indirect reciprocity approach with different $\mu$ at $p_{e}=0.01$.}
\end{center}
\end{figure}

\begin{figure}[h!]
\centering
\begin{tabular}{c}
\includegraphics[width=8.6cm]{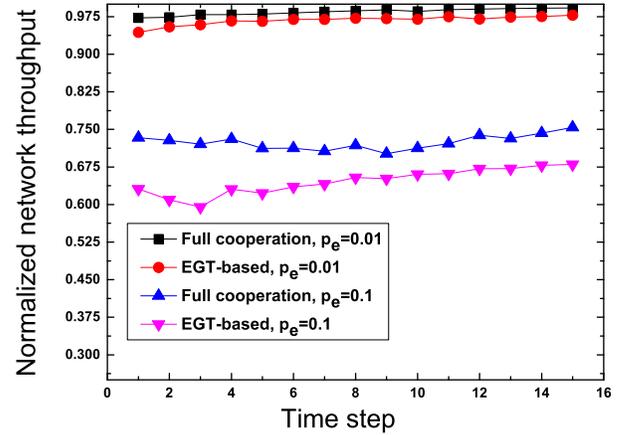}
\end{tabular}
\begin{center}
\caption{Normalized network throughput for different channel loss probabilities based on the EGT approach in the SS at $\mu=0.01$.}
\end{center}
\end{figure}

To further evaluate our proposed strategies in the SS, we consider the network performance assuming every hop on a data route is independent.
Denote $1$ as the state that all the packets are successfully delivered from a source to a destination.
In Fig. 9, we present the normalized network throughput with different $p_{e}$ when $\mu=0.01$ in the SS.
The throughput difference between the EGT-based indirect reciprocity approach and ``Full cooperation" becomes larger, when channel loss probability ($p_{e}=0.1$) is large.
In Fig. 10, we present the normalized network throughput with different $\mu$ when $p_{e}=0.01$ in the SS.
It is shown that the throughput of the EGT-based indirect reciprocity approach and ``Full cooperation" decreases with the increase of $\mu$.
Moreover, the throughput difference between the EGT-based indirect reciprocity approach and ``Full cooperation" also becomes larger, when $\mu$ ($\mu=0.1$) is large.
Both figures show that with a small channel noise ($p_{e}=0.01$, $\mu=0.01$), our proposed EGT-based indirect reciprocity approach reaches almost the same throughput as that of the fully cooperative strategy.


In summary, decreasing $p_{e}$ and $\mu$ improves the performance of our proposed strategy both in the USS and the SS.
The proposed EGT-based indirect reciprocity approach significantly promotes cooperation of packet forwarding in unreliable MANETs.
Particularly, when the channel noise is small, the performance evaluation of our proposed strategy is much closer to the performance evaluation of the full cooperative strategy.
Besides, all the figures show that the performance evaluation of the average payoff and the normalized network throughput in the SS is advantageous over the performance evaluation in the USS, which suggests that the proposed strategy in the SS outperforms in the USS.

\subsection{Evolutionary dynamics}

\begin{figure}[h!]
\centering
\begin{tabular}{c}
\includegraphics[width=8.5cm]{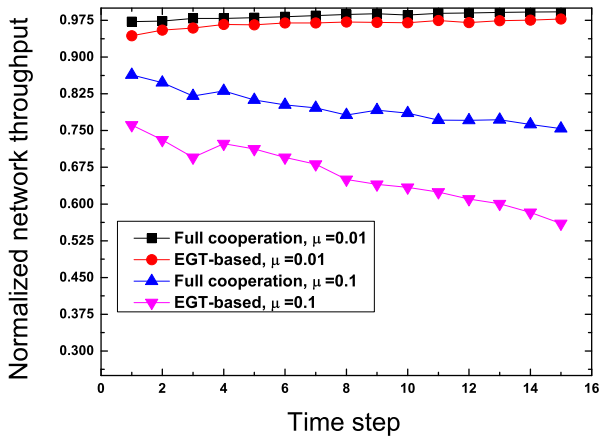}
\end{tabular}
\begin{center}
\caption{Normalized network throughput for different reputation updating errors based on EGT approach in the SS at $p_{e}=0.01$.}
\end{center}
\end{figure}

To study how cooperative strategies evolve in the packet forwarding game, we plot the cooperation frequency.
The cooperative strategies include $FF$ and $FD$.
When the player adopts $FD$, it has probability $1-\mu$ to meet a good reputation player.
Therefore, the frequency of cooperation $x_{f}=x_{1}+x_{2}(1-\mu)$.

\begin{figure}[h]
\centering
\begin{tabular}{c}
\includegraphics[width=7.3cm]{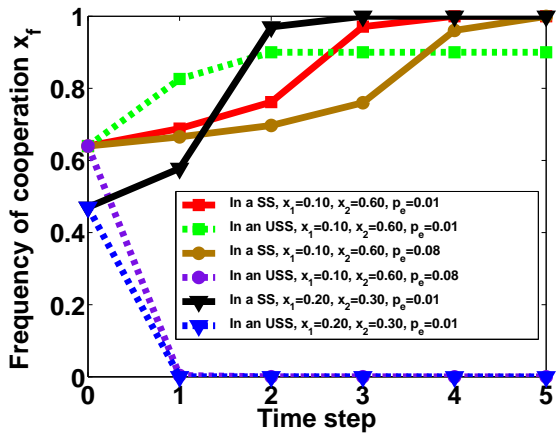}
\end{tabular}
\begin{center}
\caption{Evolutionary dynamics of cooperative strategy versus initial frequency of strategies and channel loss probability.}
\end{center}
\end{figure}

In Fig. 11, we choose three arrays of data (\textit{(1)} $x_{1}=0.1$, $x_{1}=0.6$, and $p_{e}=0.01$; \textit{(2)} $x_{1}=0.1$, $x_{1}=0.6$, and $p_{e}=0.08$; \textit{(3)} $x_{1}=0.2$, $x_{1}=0.3$, and $p_{e}=0.01$) to illustrate the strategy dynamics with different initial frequencies of strategy and channel loss probability, and set $\mu=0.1$.
It shows that the frequency of cooperation in the SS is higher than that in the USS with the same parameters and initial frequency.
When the parameters in the USS satisfy the condition in Theorem 1, the entire population converges to $FD$.
When the parameters in the SS satisfy the condition \textit{(I)} in Theorem 2, the entire population converges to $FF$.
It also suggests that the more reliable the channel is, the higher level of cooperation can be enforced.
Besides, a larger initial frequency of cooperative strategy ($x_{1}+x_{2}$) also results in a higher level of cooperation.

\section{Conclusion and Discussion}
MANETs require all nodes in a network to cooperatively conduct a task,
where the lack of a single authority and the limited battery resources are likely to lead to a noncooperative behavior at the level of packet forwarding.
In this paper, we have proposed an evolutionary game theoretic solution to enforce cooperation not only in the USS, but also in the SS.
Based on the indirect reciprocity mechanism, we have theoretically analyzed the evolutionary dynamics of cooperative strategies,
and derived the approximate threshold of benefit-to-cost ratio to guarantee the convergence of cooperation.
From the simulation results, we find that the indirect reciprocity mechanism promotes the evolution of cooperation in both the USS and the SS,
and the state of system converges to the cooperative ESS ($FD$ or $FF$).
In particular, the proposed strategy in the SS outperforms that in the USS.

Game theory has been applied to analyze an integrated model of transmission losses, buffer overflows, packet forwarding and routing information dissemination in self-organized MANETs.
In this paper, we start the analysis of the packet forwarding problem by considering a simpler game between two nodes that take turns to send their packets,
in such a way that each node requires the retransmission services of the other, as shown in Fig. 1.
Although this two-node scenario is a simplified model, we build an analytically tractable, non-cooperative game with incomplete information,
the Forwarding Dilemma (FD).
The analysis method we devised show its superiority over the classical prisoner dilemma of reputation model of MANETs,
due to the evolutionarily stable strategies based on indirect reciprocity is effective and robust against packet loss and imperfect estimation of reputation.
Besides, our analysis method shed light on the study of the multi-hop packet forwarding model in MANETs.

\newpage

\section*{Appendix}

\subsection{Notations}
\begin{table}[h!]
\caption{Notations in the paper}
\begin{tabular}{|c|c|}
\hline
\qquad Notations & Physical Meanings\\
\hline
$\mathbb{A}$ & strategy set of packet forwarding game\\
\hline
$a_{i}$ & strategy of player $i$ ($a_{i}\in\mathbb{A}$)\\
\hline
$\mathbb{\tilde{A}}$ & strategy set within the framework of indirect reciprocity\\
\hline
$\tilde{a}_{i}$ & strategy of player $i$ ($\tilde{a}_{i}\in\mathbb{\tilde{A}}$)\\
\hline
$\Theta$ & set of observed signal\\
\hline
$\theta_{i}$ & observed signal of player $i$\\
\hline
$b$ & gain of a player as a provider\\
\hline
$c$ & cost of a player as a relay \\
\hline
$p_{e}$ & channel loss probability \\
\hline
$\mu$ & reputation updating error \\
\hline
$x_{m}$ & frequency of strategy $m$\\
\hline
$r_{m}$ & frequency of strategy $m$ with good reputation\\
\hline
$r$ & sum frequency of players with good reputation\\
\hline
$N$ & Number of players in network\\
\hline
$L$ & degree of network \\
\hline
$k_{XY}$ & broken probability of $XY$-type link \\
\hline
\end{tabular}
\end{table}

\subsection{Proof of Lemma 1}
{\it Proof:}
With the fast reputation updating mechanism, the time scale of players' strategy updating is much slower than reputation updating.
Thus, the frequency of strategies is fixed during the reputation process.
In this case, a $FF$ player has $\frac{1}{2}$ chance to be a provider,
his reputation does not change and remains as the current frequency $r_{1}$.
On the other hand, this $FF$ player has $\frac{1}{2}$ chance to be a relay, and takes cooperation strategy no matter what reputation the provider has.
Due to the updating error, he gets a good reputation with probability $1-\mu$ and a bad reputation with probability $\mu$.
According to this recursive process, we derive that the new frequency of players with good reputation in $FF$ players is $\tilde{r}_{1}=\frac{1}{2}r_{1}+\frac{1}{2}(1-\mu)$.
Similarly, we get $\tilde{r}_{2}=\frac{1}{2}r_{2}+\frac{1}{2}(1-\mu)$ and $\tilde{r}_{3}=\frac{1}{2}r_{3}+\frac{1}{2}[(1-\mu)(1-r)+r\mu]$, respectively.
Therefore, we easily obtain the stable reputation frequency of each strategy as Eq. $\eqref{eq:50}$.

\section*{Acknowledgements}
The authors were grateful to Yue-Dong Xu, Zhong-Hua Yang and Bin Wu for their helpful discussions and valuable suggestions.

\end{document}